\documentclass[sigconf, screen, nonacm]{acmart}

\copyrightyear{2026}
\acmYear{2026}
\setcopyright{cc}
\setcctype{by}
\acmConference[ICAIF '26]{7th ACM International Conference on AI in Finance}{November 14--17, 2026}{Milan, AA, Italy}
\acmBooktitle{7th ACM International Conference on AI in Finance (ICAIF '26), November 14--17, 2026, Milan, AA, Italy}
\acmDOI{10.1145/3858063.3858845}
\acmISBN{979-8-4007-3087-0/2026/11}

\begin{document}

\title{Arbitrage-Aware Probabilistic Multi-Step Forecasting of Implied Volatility Surfaces}
\subtitle{Joint Generation of Surface-Return Trajectories Using Latent Diffusion}

\author{Dominik Manuel Buchegger}
\correspondingauthor
\orcid{0009-0000-9417-2361}
\affiliation{%
  \institution{University of St.Gallen}
  \city{St.Gallen}
  \country{Switzerland}}
\email{me@dominikbuchegger.com}
\author{Lukas Gonon}
\orcid{0000-0003-3367-2455}
\affiliation{%
  \institution{University of St.Gallen}
  \city{St.Gallen}
  \country{Switzerland}}
\email{lukas.gonon@unisg.ch}
  
\begin{abstract}
  Implied volatility surfaces summarise the option market and are central to many financial applications. Probabilistic forecasting of their future evolution requires modelling two-dimensional geometry, temporal dependence, and predictive uncertainty while preserving economic admissibility. We propose a conditional latent diffusion framework for generating joint 30-step trajectories of implied volatility surfaces and underlying returns. An arbitrage-aware autoencoder learns a surface representation, while diffusion captures the evolution. Evaluated on SPX surfaces, the framework generates realistic probabilistic multi-step scenarios. Its predictive mean also outperforms persistence beyond the shortest horizons.
\end{abstract}

\ccsdesc[500]{Computing methodologies~Machine learning}
\keywords{implied volatility surfaces, latent diffusion models, probabilistic forecasting, multi-step forecasting, time series forecasting, joint return-volatility generation, static arbitrage, option markets}

\maketitle

\section{Introduction}
\label{sec:introduction}

The \emph{implied volatility surface (IVS)} summarises option-implied uncertainty across moneyness and maturity and is a core input to pricing, hedging, and risk management. While IVS shapes vary across assets and time, three properties frame the modelling problem. First, an admissible IVS must satisfy shape restrictions excluding \emph{static arbitrage}. Second, a probabilistic forecast of future \emph{IVS trajectories}, the IVS evolution over multiple days, should capture the observed temporal and cross-sectional dependence structure. Third, the IVS is rather \emph{persistent}, so its current state is already a strong prediction of its future evolution. A useful model must demonstrate value beyond carrying the most recent surface forward. Together, these properties make point-forecast accuracy insufficient on its own. A model copying the current IVS can yield low errors without learning much about the conditional law of future surface movements. A useful forecasting model must instead respect the economic structure of option markets, generate informative trajectories with realistic uncertainty, and demonstrate predictive value relative to persistence rather than merely visual plausibility or realistic unconditional sampling.

Existing work does not jointly address these requirements. In particular, probabilistic forecasting of complete IVS trajectories has received almost no attention. We address this gap with a two-stage \emph{latent diffusion} model. An \emph{autoencoder} compresses daily surfaces into latent codes, while its decoder is regularised towards static no-arbitrage consistency, encouraging admissibility directly during reconstruction rather than through a parametric family or post-processing. On these latent codes, a conditional \emph{diffusion} model generates the next \(30\) days of surfaces jointly with daily equity returns in one non-autoregressive pass, conditioned on recent surface history. This separation assigns representation to the autoencoder and the conditional law over future paths to the diffusion model.

Our contribution is twofold. First, we provide a transparent, reproducible protocol for constructing IVS forecasting datasets from \emph{OptionMetrics} data on \emph{WRDS}, allowing licensed users to recreate a common evaluation setting. Second, we develop a latent diffusion model for conditional, multi-step generation of joint IVS--return trajectories and evaluate it systematically against persistence. Code is available on \href{https://github.com/DomBBB/implied-volatility-trajectories/tree/main/ICAIF26}{GitHub}\footnote{\url{https://github.com/DomBBB/implied-volatility-trajectories/tree/main/ICAIF26}}. Together, the data protocol, model, and persistence baseline constitute a first step towards a reproducible benchmark for a forecasting problem that currently lacks one.

The results support the model primarily as a probabilistic scenario generator rather than a one-day point forecaster. Generated surfaces preserve the dominant smile and term-structure geometry, reproduce the principal empirical modes of variation in level, skew, and curvature, and capture negative return--volatility dependence particularly well at weekly-to-monthly horizons. Generated trajectories improve probabilistic forecast scores over persistence-based benchmarks at medium and longer horizons, with negligible detected static-arbitrage violations. The main limitation is underdispersion: predictive intervals widen with horizon but remain too narrow, indicating that the model captures movement patterns and dependence more accurately than their amplitude. Beyond this distributional performance, the predictive mean also outperforms persistence from approximately weekly horizons onward, although persistence remains superior at one day. Gains are strongest at short maturities and in the wings, while the stable long-maturity centre remains difficult to improve upon.

\section{Related Work}
\label{cp:related-work}

\paragraph{Representation and admissibility}
Implied volatility surfaces are empirically low-dimensional, with level, skew, and curvature explaining most daily variation \cite{Cont02}. Useful representations must also respect (static) no-arbitrage restrictions \cite{Roper10}. Learnt models address this either by operating within an admissible parametric family \cite{NingJaimungal23} or by combining a flexible representation with soft arbitrage penalties \cite{Ackerer20,Wiedemann24}. SANOS provides a recent nonparametric alternative that is arbitrage-free by construction \cite{BuehlerEtAl26SANOS}.

\paragraph{Dynamic surface forecasting}
Most prior machine-learning approaches produce deterministic single-step forecasts. \citet{ZhangLiZhang23} compress the surface, forecast its low-dimensional features, and reconstruct. \citet{CaoChenHull20} regress expected IVS changes on returns, moneyness, maturity, and VIX. Neither yields predictive uncertainty. Generative work began without dynamics. VAEs learn latent spaces from which realistic surfaces can be decoded \cite{BergeronFungHull22} or learn the parameter space of an SDE family to obtain admissibility by design \cite{NingJaimungal23}, an approach extended to dynamic neural-SDE market models \cite{Cohen23}. Dynamics also appear in one-step generative approaches: \citet{VuleticCont25} generate surfaces jointly with the underlying (controlling arbitrage through a smoothness penalty and reweighting) and \citet{ChenHullPoulos25} generate conditional surfaces from historical sequences using CVAE--LSTMs.

Existing multi-step methods remain mostly deterministic, e.g., ConvLSTMs applied to discretised IV values without arbitrage constraints \cite{MedvedevWang22}, SVI coefficient dynamics with admissibility resting on the parametric family \cite{Bloch21}, and functional autoregression on surface time series \cite{ChenGrithLai25}. Our closest comparator is \citet{ChoudharyJaimungalBergeron24}, who combine functional PCA with neural SDEs to generate surfaces jointly with price paths and assess admissibility empirically.

DYSANOS by \citet{BuehlerEtAl26DYSANOS} takes a complementary route to dynamic generation by representing arbitrage-free (SANOS \cite{BuehlerEtAl26SANOS}) surfaces through a low-dimensional state and modelling its dynamics. Its objective is generative market simulation rather than conditional forecasting, but it provides a related alternative for combining low-dimensional representations with learnt dynamics.

\paragraph{Latent diffusion}
Existing diffusion applications to IVSs focus on surface completion and generation of synthetic surfaces \cite{Ma23,Hui24}, and, closest to us, one-day-ahead forecasting with an SNR-weighted arbitrage penalty \cite{JinAgarwal25}. However, these applications operate directly on surfaces and are limited to static or single-step settings.

Concurrent work by \citet{BrooksBajalica26} adopts a similar latent generative architecture to ours, replacing diffusion with flow matching, but focuses on unconditional generation of individual surfaces.

\paragraph{Persistence rarely confronted}
A rarely confronted difficulty runs through the forecasting literature: daily IVSs are highly persistent. Consequently, low errors and visually plausible forecasts may reflect little more than copying the latest surface. This risk is especially acute for models conditioned primarily on the current surface. Yet few forecasting studies thoroughly evaluate against a persistence baseline, and fewer beat it convincingly. 

\section{Theoretical Background}
\label{sec:background}

\paragraph{Options}
\emph{Options} grant their holder the right, but not the obligation, to buy (a \emph{call}) or sell (a \emph{put}) an underlying in the future on fixed terms, notably transaction price (\emph{strike} \(K\)) and remaining contract lifetime (\emph{time to maturity} \(\tau\)). Options are traded for different reasons, including speculation and hedging. Consequently, there needs to be a way to determine option values, which requires a model for how uncertainty in the underlying's price up to maturity propagates into value. The \emph{Black--Scholes--Merton (BSM)} model \cite{BlackScholes73,Merton73} supplies one under a stylised set of assumptions. Of the inputs to these call (\(C_t^{BSM}\)) and put (\(P_t^{BSM}\)) formulas, all are contractual, observed, or estimated -- except volatility \(\sigma\). Volatility governs the uncertainty of the underlying over the option's remaining life and is unobservable ex ante. However, with prices \(C_t, P_t\) observed on the market, the \emph{(BSM) implied volatility (IV)} \(\hat{\sigma}\) is defined as the volatility value that reproduces the observed market price, e.g., \(C_t^{BSM}(\hat{\sigma})=C_t\).

Under BSM, \(\sigma\) is constant across all options on a given underlying, so each option should imply the same IV -- the market's assessment of future uncertainty over the option's life. Empirically, however, IV varies with both strike and maturity. The resulting function, \(\Sigma : (K,\tau) \mapsto \hat{\sigma}(K,\tau)\), is the \emph{implied volatility surface (IVS)}. Its non-flatness reflects the failure of BSM's constant-volatility assumption to explain the complete option-price cross-section. Nevertheless, BSM IV remains the standard convention in which option prices are quoted and compared across strikes, maturities, and dates.\cite{Gatheral06}

\paragraph{Surface construction and static arbitrage}
The IVS is continuous, whereas markets provide only a finite irregular collection of quotes. These quotes are additionally affected by bid--ask spreads, non-synchronous recording, and occasional errors. Constructing a surface therefore requires interpolation, smoothing, and extrapolation across unobserved strike--maturity combinations.\cite{Fengler12}

A central requirement is the absence of \emph{static arbitrage}. Unlike dynamic arbitrage, static arbitrage can be identified from the contemporaneous option-price cross-section without specifying a stochastic model for the underlying \cite{Carr05}. It therefore provides a model-independent notion of internal consistency for a fitted or generated surface, and translates into shape constraints on the IVS, specifically on the \emph{time-scaled IVS (t--IVS)} in \emph{log--forward--moneyness} reparametrisation, \(\Xi(m,\tau) = \Sigma\!\left(m,\tau\right)\sqrt{\tau}\;\). The six model-independent no-static--arbitrage conditions\footnote{Stated for zero rates and dividends; the functional forms are unchanged for non-zero parameters in log--forward--moneyness \cite{Ackerer20}.} are
\[
\begin{aligned}
\makebox[3.2cm][l]{\text{C1) Smoothness}}\qquad
& \forall \tau > 0: \Xi(\cdot,\tau) \text{ twice differentiable} \\
\makebox[3.2cm][l]{\text{C2) Positivity}}\qquad
& \forall \tau > 0,\ \forall m: \Xi > 0 \\
\makebox[3.2cm][l]{\text{C3) Durrleman's condition}}\qquad
& \forall \tau > 0,\ \forall m: 
\Bigl(1-\tfrac{m\partial_m \Xi}{\Xi}\Bigr)^2
- \\
\makebox[3.2cm][l]{}\qquad
& \tfrac14 \Xi^2(\partial_m \Xi)^2
+\Xi \partial_{mm}^2 \Xi \geq 0  \\
\makebox[3.2cm][l]{\text{C4) Monotonicity in \(\tau\)}}\qquad
& \forall m: \Xi(m,\cdot)\ \text{non-decreasing} \\
\makebox[3.2cm][l]{\text{C5) Large moneyness}}\qquad
& \forall \tau>0 : \lim_{m\to\infty} d_+\bigl(m,\Xi\bigr) = -\infty \\
\makebox[3.2cm][l]{\text{C6) Value at maturity}}\qquad
& \forall m: \Xi(m, 0) = 0
\end{aligned}
\]

Condition C3 excludes \emph{butterfly arbitrage}, condition C4 excludes \emph{calendar--spread arbitrage}, and condition C5 concerns the asymptotic large-moneyness behaviour \cite{Roper10}. In the bounded empirical domain studied in our work, only C1--C4 can be evaluated, since C5 and C6 concern behaviour outside its bounds.
\section{Methodology}
\label{sec:methodology}

\subsection{Data}
\label{sec:methodology:data}
We construct daily IVSs from OptionMetrics \emph{IvyDB US} end-of-day SPX option quotes obtained through WRDS \cite{OptionMetrics24,WRDS26}. The sample covers January \(2000\) to August \(2025\). We apply \emph{operator deep smoothing} \cite{Wiedemann24} for each date: a \emph{\href{https://github.com/DomBBB/implied-volatility-trajectories/blob/main/ICAIF26/operator-deep-smoothing/volatility_smoothing/train/store/9448705/checkpoints/checkpoint_final.pt}{pretrained} \footnote{\url{https://github.com/DomBBB/implied-volatility-trajectories/blob/main/ICAIF26/operator-deep-smoothing/volatility_smoothing/train/store/9448705/checkpoints/checkpoint_final.pt}} graph neural operator (GNO)} maps the irregular (BSM) IV quote set to a smooth, essentially static--arbitrage-free surface on prescribed coordinates. Our initial coordinate grid is 
\(\mathcal{G}_{\mathrm{orig}} = \mathcal{M}\times\mathcal{T},\mathcal{M}=\{-1.5,-1.4,\dots,0.5\},\mathcal{T}=\left\{{1}/{12},\dots,{12}/{12}\right\}\). We retain \(170\) coordinates that lie within the domain supported by the pretrained operator (\((\sqrt{\tau}\in [0.01,1],{m}/{\sqrt{\tau}}\in [-1.5,0.5])\)), yielding identical grid availability on every date,
\[
    \mathcal{G}_{\mathrm{obs}} 
    =
    \{(m,\tau)\in\mathcal{G}_{\mathrm{orig}} \mid \omega(m,\tau)=1\}
    \subseteq
    \mathcal{G}_{\mathrm{orig}},
\]
where \(\omega(m,\tau)\in\{0,1\}\) denotes the observation indicator on \(\mathcal{G}_{\mathrm{orig}}\). 

To provide the diffusion model with information about recent underlying dynamics, the time-scaled surface (t--IVS) history is augmented with SPX log returns and rolling summaries of return levels and return volatility. The complete surface-construction pipeline is provided in the accompanying code repository. 

The resulting dataset contains \(6429\) daily t--IVSs and associated conditioning variables. We split it chronologically into \(5511\) \emph{training} observations through \(2021\), \(251\) \emph{validation} observations from \(2022\), and \(667\) \emph{test} observations from \(2023\) onward.

\subsection{Latent Surface Representation}
\label{sec:method:autoencoder}

First, we transform each daily t--IVS as \(y_i := \log \Xi_i\) and standardise it. Modelling \(y^{\mathrm{norm}}_i\) would require the diffusion model to learn dynamics in a high-dimensional and strongly constrained space. We therefore compress each \(y^{\mathrm{norm}}_i\) into an eight-dimensional latent state using a deterministic, RAE-inspired \cite{Ghosh20} autoencoder. Recent work on the role of traditionally used variational autoencoders in latent diffusion suggests that diffusion may benefit from different autoencoder designs, including deterministic variants with additional architectural constraints and regularisations.

Architectural choices for the autoencoder matter because they affect reconstruction and the latent space. For our task, the autoencoder must construct a latent that preserves enough information to reconstruct IVSs while being sufficiently regularised so that changes in latent space can be decoded into admissible surfaces. We use an \emph{RAE-inspired} architecture that is deterministic and shapes the latent space through explicit and implicit regularisation.

\begin{figure}[!htpb]
    \centering
    \includegraphics[width=\columnwidth]{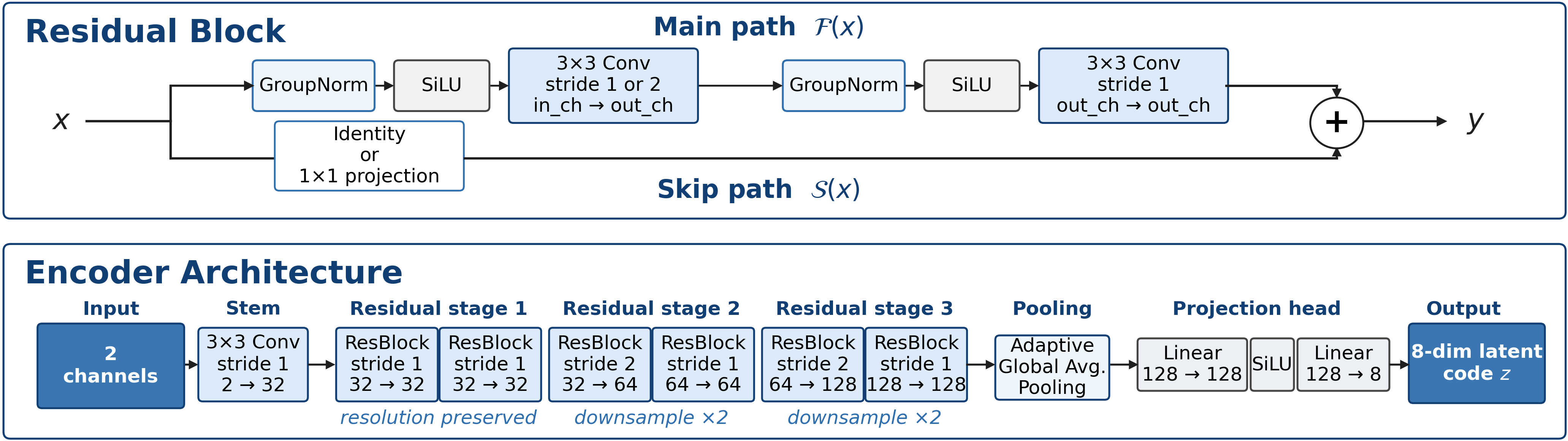}
    \Description{Two stacked architecture diagrams. The upper diagram shows a residual block in which an input splits into a main path and a skip path. The main path applies Group Normalisation, SiLU activation, a three-by-three convolution, another normalisation and activation, and a second convolution; the skip path is either the identity or a one-by-one projection, and both paths are added. The lower diagram shows the full encoder: a two-channel surface and observation-mask input passes through an initial convolution, three residual stages with two spatial downsamplings, adaptive global average pooling, and a linear projection head, producing an eight-dimensional latent code.}
    \caption[Encoder architecture]{Encoder: mapping the IVS to a latent state.}
    \label{fig:ae_encoder}
\end{figure}

\paragraph{Encoder}
The encoder is a residual convolutional network \cite{HeZhangRenSun16} illustrated in \autoref{fig:ae_encoder}. It receives the normalised log surface \(y^{\mathrm{norm}}_i\) and the common observation mask as two input channels and outputs an eight-dimensional latent code \(z_i\).

\begin{figure}[!htpb]
    \centering
    \includegraphics[width=\columnwidth]{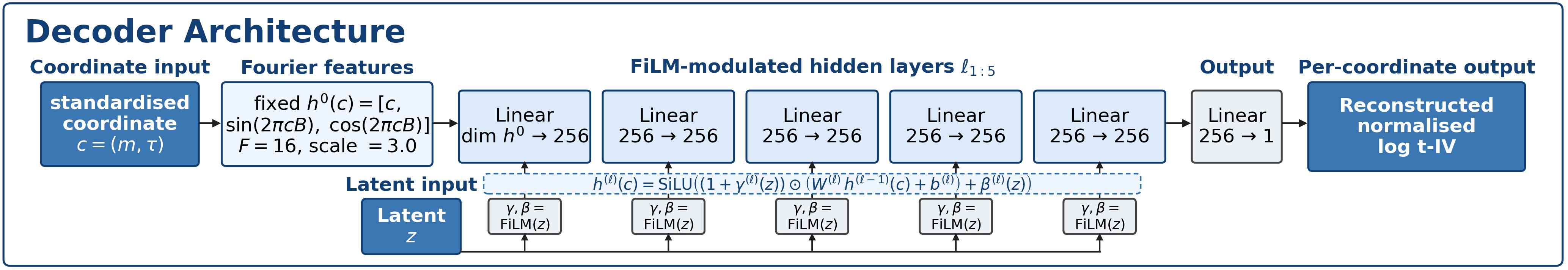}
    \Description{A left-to-right diagram of the coordinate-based decoder. A standardised maturity--moneyness coordinate is converted to fixed Fourier features and passed through five fully connected hidden layers of width 256. An eight-dimensional latent code enters below and produces feature-wise scale and shift parameters for every hidden layer through FiLM modules. A final linear layer maps each conditioned hidden representation to one reconstructed normalised log time-scaled implied volatility value at the queried coordinate.}
    \caption[Decoder architecture]{Decoder: mapping m--\(\tau\) coordinates to IV values.}
    \label{fig:ae_decoder}
\end{figure}

\paragraph{Decoder}
The decoder is a coordinate-based neural field \cite{Xie22}, illustrated in \autoref{fig:ae_decoder}. It reconstructs the normalised log--t--IVS \(\hat y_i^{\mathrm{norm}}\) continuously for each queried coordinate \(c=(m,\tau)\).

\paragraph{Training objective}
We write the terms for a minibatch \(\mathcal{B}\). \\ The \emph{reconstruction term} is the masked mean squared error,

\[
    \mathcal{L}_{\mathrm{rec}}
    =
    \frac{
    \sum_{i\in\mathcal{B}}
    \sum_{(m,\tau)\in\mathcal{G}_{\mathrm{orig}}}
    \omega(m,\tau)\,
    \bigl(
    \hat y_i^{\mathrm{norm}}(m,\tau)
    -
    y_i^{\mathrm{norm}}(m,\tau)
    \bigr)^2
    }{
    \sum_{i\in\mathcal{B}}
    \sum_{(m,\tau)\in\mathcal{G}_{\mathrm{orig}}}
    \omega(m,\tau)
    }.
\]
The \emph{latent regularisation term} regularises the representation directly,
\[
    \mathcal{L}_{\ell2}
    =
    \frac{1}{|\mathcal{B}|\,\dim(z_i)}
    \sum_{i\in\mathcal{B}}
    \|z_i\|_2^2.
\]
The two \emph{no-static--arbitrage terms} guide the model. Smooth operations in the architecture make the decoder twice differentiable (C1); the logarithmic surface parametrisation guarantees positivity (C2). Butterfly (C3) and calendar--spread (C4) conditions are evaluated on \(\mathcal{G}_{\mathrm{dense}}\), a denser set that is useful because the decoder represents a continuous coordinate-based surface.\footnote{
\( 
    \mathcal{G}_{\mathrm{obs}}
    \subsetneq
    \mathcal{G}_{\mathrm{dense}}
    =
    \Bigl\{(m,\tau)\in[-1.5,0.5]_{0.02} \;\tilde{\times}\; [\tfrac{1}{12},1]_{0.02}\Bigr\};
\)
\([\cdot]_{0.02}\) denotes the discretised interval with step size \(0.02\); \(\tilde{\times}\) indicates restriction to observation bounds.} In log t--IV parametrisation, we have \(0 \leq \Bigl(1-m y_m\Bigr)^2 -\tfrac14 e^{4y} y_m^2 +e^{2y}\Bigl(y_m^2 + y_{mm}\Bigr) =: l_{\mathrm{but}}\) and \(0 \leq y_\tau =: l_{\mathrm{cal}}\). Negative values of \(l_{\mathrm{but}}\) and \(l_{\mathrm{cal}}\) correspond to violations and are therefore penalised as follows, for \(\mathrm{cond} \in \{\mathrm{but},\mathrm{cal}\}\):
\[
    \mathcal{L}_{\mathrm{cond}}
    =
    \frac{1}{|\mathcal{B}|\,|\mathcal{G}_{\mathrm{dense}}|}
    \sum_{i\in\mathcal{B}}
    \sum_{(m,\tau)\in\mathcal{G}_{\mathrm{dense}}}
    \left[
    \max\left\{0,-l_{\mathrm{cond},i}(m,\tau)\right\}
    \right]^2.
\]

\paragraph{Optimisation}
The autoencoder is trained by minimising \(\mathcal{L}=\mathcal{L}_{\mathrm{rec}} +\lambda_{\ell2}\mathcal{L}_{\ell2} +\lambda_{\mathrm{but}}\mathcal{L}_{\mathrm{but}} +\lambda_{\mathrm{cal}}\mathcal{L}_{\mathrm{cal}}\), with \(\lambda_{\ell2}=10^{-4}, \lambda_{\mathrm{but}}=10^{-3}, \lambda_{\mathrm{cal}}=10^{-4}\). Architectural choices are selected on the validation set. The final model is trained on the combined training and validation sample for \(200\) epochs. Weight decay acts as parameter-space regularisation, providing an explicit regularisation mechanism of the RAE \cite{Ghosh20} framework, complemented by implicit regularisation associated with CNNs and gradient-based optimisation.

\subsection{Conditional Latent Trajectory Diffusion}
\label{sec:methodology:diffusion}

The autoencoder converts the daily surfaces into latent states \(z_i\) that are then used to construct trajectories. We model their future evolution with a conditional latent diffusion model, avoiding direct generation in the high-dimensional surface space \cite{Ho20,Rombach22}.

\begin{figure*}
    \centering
    \includegraphics[width=\textwidth]{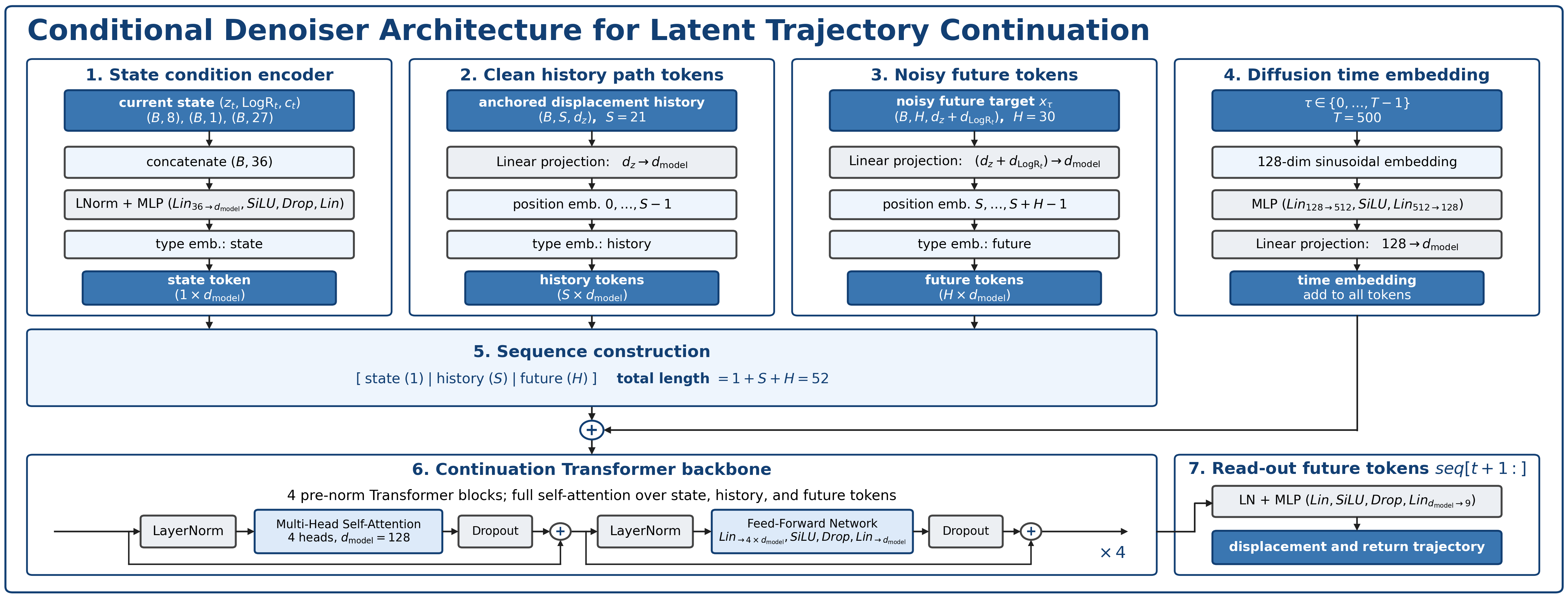}
    \Description{A seven-stage architecture diagram for joint latent-surface and return continuation. Four upper branches construct a current-state token, 21 clean history tokens, 30 noisy future tokens, and a diffusion-time embedding. The state contains the current latent, return, and conditioning variables; history tokens contain anchored latent displacements; future tokens contain noisy latent displacements and returns. These are combined into a sequence of length 52, augmented by the diffusion-time embedding, and processed by four pre-normalised Transformer blocks with self-attention and feed-forward layers. The 30 future-token outputs are read out through a multilayer perceptron to predict the denoising targets for the latent-displacement and return trajectory.}
    \caption[Conditional latent trajectory denoiser]{Transformer denoiser for joint latent surface and return continuation.}
    \label{fig:diffusion}
\end{figure*}

At forecast origin \(t\), the diffusion model observes \(S=21\) trading days and jointly generates the next \(H=30\) latent surface states and daily returns. The time-series diffusion literature provides different strategies for achieving temporal and conditional coherence, e.g., \cite{Rasul21,ShenKwok23,Kollovieh23,Feng24}. We formulate our task as a one-sided inpainting problem, asking the model to plausibly extend observed trajectories.

\paragraph{Trajectory representation}
We represent each latent by \emph{anchored displacements},
\(\Delta^{\mathrm{anc}}z_s = z_s-z_{t_0}\), with \(t_0=t-S+1\). Using a common anchor places observed and future trajectories in the same coordinate system  while separating evolution from absolute latent levels. Persistence of the latest observed state corresponds to future anchored displacements remaining at their value at the forecast origin. This aligns the training target with the object of interest: conditional departures from persisting the latest observed state.

The clean diffusion target \(x_0 = \left[\Delta^{\mathrm{anc}} z^{\mathrm{norm}}_{t+1:t+30},\mathrm{LogR}_{t+1:t+30}^{\mathrm{norm}}\right]\) is generated conditional on \(\mathcal{C}_t = \bigl(\Delta^{\mathrm{anc}} z^{\mathrm{norm}}_{t-20:t}, z^{\mathrm{norm}}_t, \mathrm{LogR}_{t}^{\mathrm{norm}},\) \(c_t^{\mathrm{norm}} \bigr)\), where \(c_t\) contains \(27\) contemporaneous surface and return-state variables. All targets and conditioning variables have been standardised (and some transformed) using training statistics only.

Forecast origins are retained only when their full \(H\)-day target is within the data split. This yields \(5461\) \emph{training}, \(221\) \emph{validation}, and \(637\) \emph{test} trajectories. After model selection, the final diffusion model is refitted on \(5712\) trajectories constructed from the combined \emph{training and validation} period.

\paragraph{Continuation denoiser}
The denoiser is a Transformer-style \cite{Vaswani17} sequence model, illustrated in \autoref{fig:diffusion}. The historical path is embedded as anchored-displacement history tokens, while the future target is embedded as future tokens containing anchored displacement and returns; positional embeddings place both on a common continuation timeline. The sequence is processed by self-attention blocks to jointly predict the denoising targets for all \(30\) future steps.

\paragraph{Diffusion training}
The forward process gradually corrupts the sequence over \(500\) diffusion steps using a linear variance schedule (\(\beta_1 = 10^{-4},\beta_{500} = 2\times10^{-2}\)). Rather than predicting \(x_0\) or \(\epsilon\) directly, we use \(v\)-parameterisation and minimise the mean squared error \(\mathcal{L}_{\mathrm{diff}} = \mathbb{E} \left[ \left\| \widehat v(x_{k},k,\mathcal{C}_t) - v_{k}\right\|_2^2\right]\) between the denoiser output and the \(v\)-target. The final model is trained for \(100\) epochs.

\paragraph{Mean-scaling gate}
The predictive mean of the diffusion distribution tended to overstate movement on the validation set. This may reflect the difficulty of learning a single rule over a long historical sample with changing market regimes and an increasingly efficient option market. We therefore freeze the trained diffusion model and subsequently fit a small horizon-specific scaling gate on the same training data. For horizon \(h\), let \(\hat z^{\mathrm{mean}}_h\) denote the mean of generated forecast samples. The gated mean is \(\hat z^{\mathrm{scaled}}_h=\hat z^{\mathrm{persistence}}+\widehat {\lambda}_h(\hat z^{\mathrm{mean}}_h-\hat z^{\mathrm{persistence}}) \). The gate is trained for \(15\) epochs by minimising the mean squared error between the scaled latent forecasts and realised latent codes. Its average fitted value is approximately \(0.80\). Instead of applying the gate to each path, we preserve the sampled variability by only shifting the predictive mean and adding back the original residuals. The gate therefore provides a horizon-dependent correction to the magnitude of predicted latent movements, based on the current market state.

\paragraph{Trajectory generation}
At inference, the complete \(30\)-day continuation is generated from Gaussian noise conditional on \(\mathcal{C}_t\). For each origin, we generate joint latent--return trajectories, \(\widehat\Delta^{\mathrm{anc}} z^{\mathrm{norm}}_{t+1:t+30}\) and \(\widehat{\mathrm{LogR}}_{t+1:t+30}^{\mathrm{norm}}\). The generated variables are de-standardised to \(\widehat\Delta^{\mathrm{anc}} z_{t+1:t+30}\) and \(\widehat{\mathrm{LogR}}_{t+1:t+30}\), and the displacements converted back to latent levels according to \(\hat z_{t+1:t+30} = \widehat\Delta^{\mathrm{anc}} z_{t+1:t+30} + z_{t_0}\). After applying the gate, each latent trajectory is passed through the decoder to obtain a corresponding \(30\)-day t--IVS trajectory.

\section{Results}
\label{sec:results}

The final pipeline is trained on the combined training and validation data and evaluated once on the held-out test set. For each forecast origin, we generate \(1000\) \(30\)-day trajectories and corresponding returns using DDPM \cite{Ho20} sampling.

\subsection{Generative Realism and Economic Admissibility}
\label{sec:results:generative-realism}

\begin{figure}[!htpb]
    \centering
    \includegraphics[width=\columnwidth]{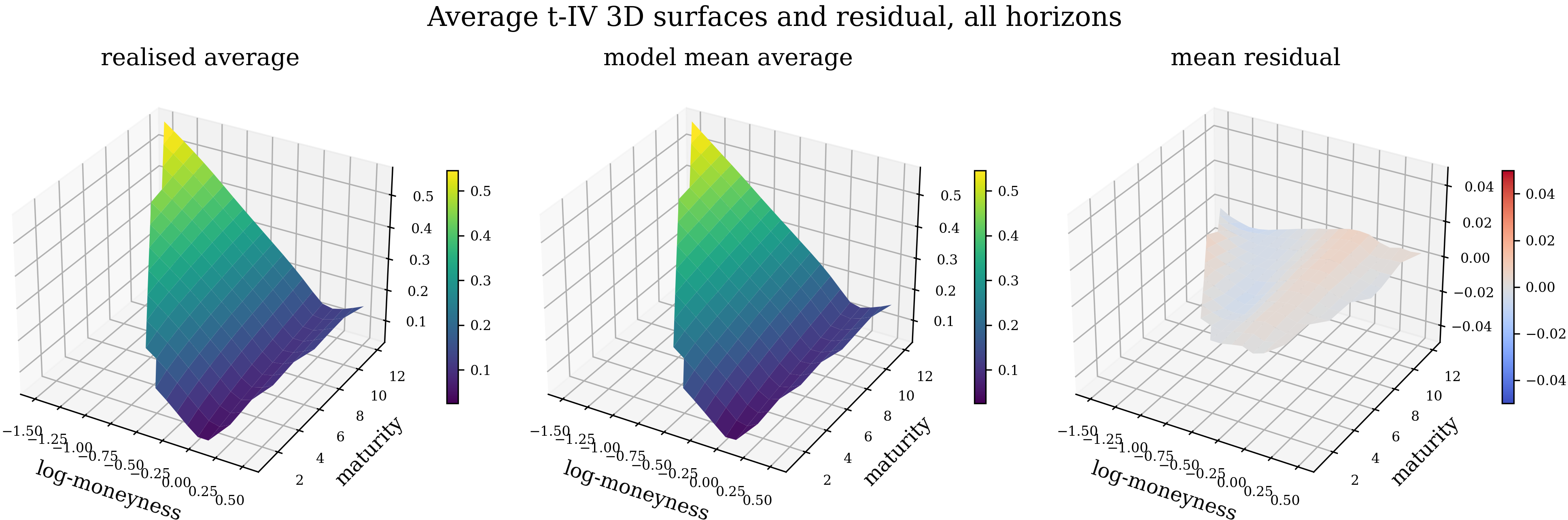}
    \Description{Three side-by-side three-dimensional surfaces over log-moneyness and maturity from one to twelve months. The realised average and model-mean average have nearly identical shapes: time-scaled implied volatility is highest for negative log-moneyness, falls towards the centre, and rises slightly on the positive-moneyness side, with levels generally increasing with maturity. The residual surface, defined as model mean minus realised value, is much flatter and centred near zero, with small negative regions on parts of the left side and small positive regions around the centre and right side.}
    \caption[Average realised and forecast t--IV surfaces]{Average realised t--IVS, model-mean forecast, and signed residual over all test origins and forecast horizons.}
    \label{fig:sec:results:generative:average-surface}
\end{figure}

\paragraph{Surface geometry}
\autoref{fig:sec:results:generative:average-surface} compares realised and model-mean t--IVSs averaged over all test origins and horizons. The model closely reproduces the dominant smile and term-structure geometry in these aggregated surfaces. Residuals are small but structured: the forecast mildly attenuates parts of the left wing and elevates the central region, particularly at longer maturities. Similar, more pronounced patterns in autoencoder-only reconstructions suggest that representation-induced smoothing contributes to the observed bias, although this comparison does not fully isolate the effects of the autoencoder and diffusion dynamics. However, a more expressive autoencoder introduces additional difficulty for the diffusion model and does not necessarily improve the full pipeline.

\begin{figure}[!htpb]
    \centering
    \includegraphics[width=\columnwidth]{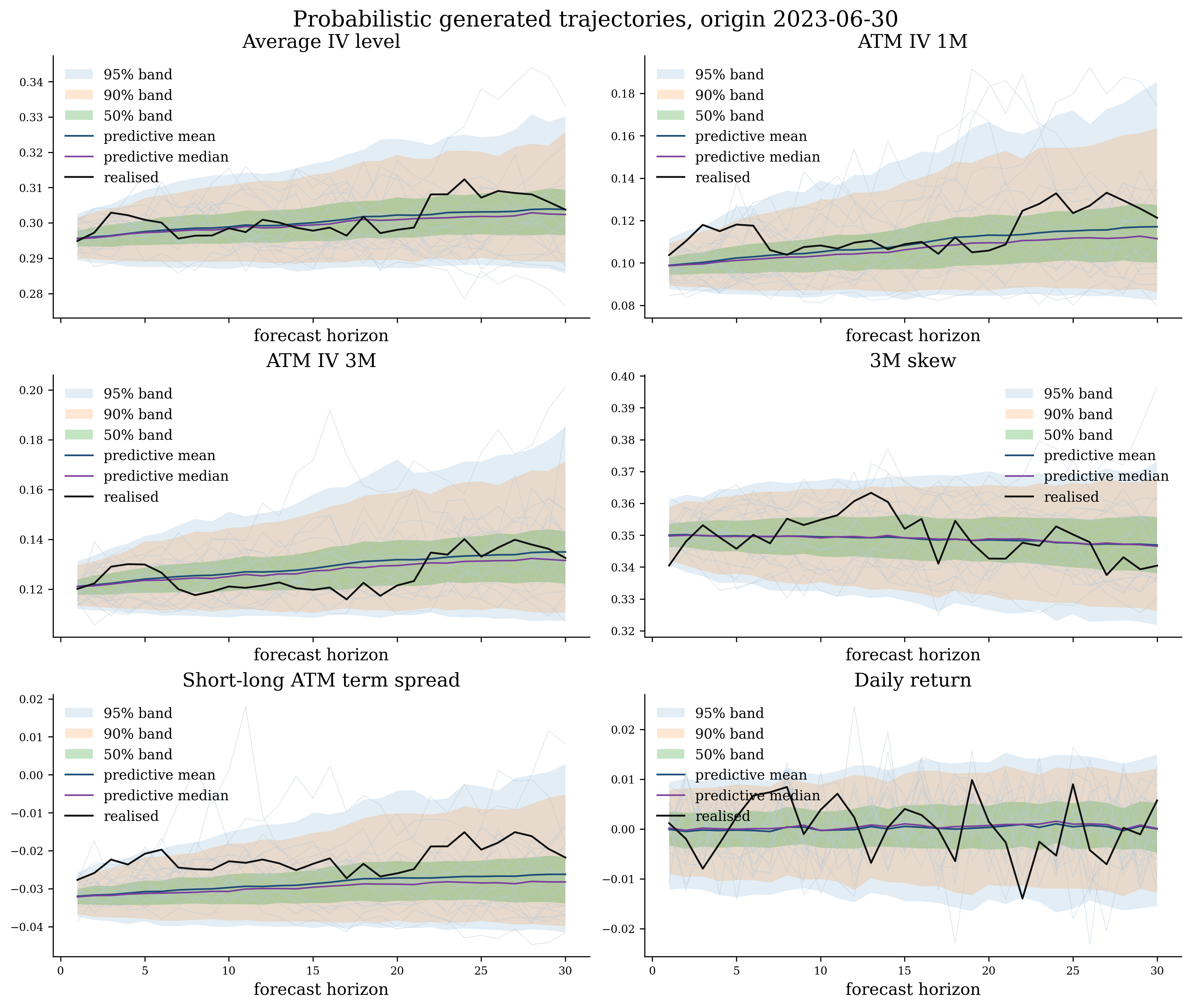}
    \Description{Six time-series panels over a 30-day forecast horizon for one forecast origin. The panels show average implied volatility, one-month at-the-money implied volatility, three-month at-the-money implied volatility, three-month skew, the short-minus-long at-the-money term spread, and daily return. Each panel contains faint individual generated paths, nested 50, 90, and 95 percent predictive bands that widen with horizon, closely overlapping predictive mean and median lines, and a black realised path. Surface-level quantities evolve smoothly around the predictive centre, while the realised skew and return paths fluctuate more sharply.}
    \caption[Illustrative generated surface and return trajectories]{Illustrative generated surface/return trajectories.}
    \label{fig:sec:results:generative:trajectories}
\end{figure}

Quantitative descriptors support these visual results. The global IV level is closely reproduced, with discrepancies mainly in local shape. These biases are small relative to variation across forecast origins, suggesting that the model preserves the dominant surface geometry while smoothing local movements. \autoref{fig:sec:results:generative:trajectories} illustrates generated paths fluctuating coherently around the predictive centre, remaining within economically plausible ranges, and producing widening intervals for level, skew, term structure, and returns.

\paragraph{Low-dimensional surface dynamics}
We fit a principal-component basis to realised log t--IVS increments in the test set and project generated increments onto that basis.\footnote{Using a common basis avoids comparing unstable principal components estimated from separate samples.}

\autoref{tab:sec:results:generative:pca-summary} shows that surface movements are low-dimensional: the first three components explain \(91.76\%\) of realised increment variance. Generated loading directions align\footnote{As a robustness check, separate PCA bases are fitted to generated paths and aligned with the realised loadings using the absolute inner product.} closely with the realised directions. The lower PC3 alignment is unsurprising given PC3's substantially smaller variance contribution.

\begin{table}[!htpb]
    \caption[Principal components of realised t--IVS increments]{Principal components of realised log t--IVS increments and generated-loading alignment.}
    \label{tab:sec:results:generative:pca-summary}
    \centering
    \footnotesize
    \begin{tabular}{lrrrr}
    \toprule
    PC & Variance share & Cum. share & Alignment & Interpretation \\
    \midrule
    PC1 & 83.95\% & 83.95\% & 0.995 & Level factor \\
    PC2 &  5.41\% & 89.35\% & 0.963 & Skew factor \\
    PC3 &  2.41\% & 91.76\% & 0.811 & Curvature factor \\
    \bottomrule
    \end{tabular}
\end{table}

\begin{table}[!htpb]
    \caption[PCA factor interpretation at \(h=21\)]{Correlations between realised-PCA scores and surface descriptors. Entries report realised/generated correlations, with generated values averaged across sample paths.}
    \label{tab:sec:results:generative:pca-interpretation}
    \centering
    \footnotesize
    \begin{tabular}{lrrrr}
        \toprule
        PC & Average IV & 3M skew & 3M curvature & Interpretation \\
        \midrule
        PC1 & \(+0.977/+0.961\) & \(-0.129/-0.261\)
            & \(-0.734/-0.698\) & Level \\
        PC2 & \(-0.117/+0.016\) & \(-0.748/-0.711\)
            & \(-0.446/-0.534\) & Skew/tilt \\
        PC3 & \(+0.086/-0.076\) & \(+0.363/+0.559\)
            & \(+0.384/+0.636\) & Curvature/bend \\
        \bottomrule
    \end{tabular}
\end{table}

\begin{figure}[!htpb]
    \centering
    \includegraphics[width=.32\columnwidth]{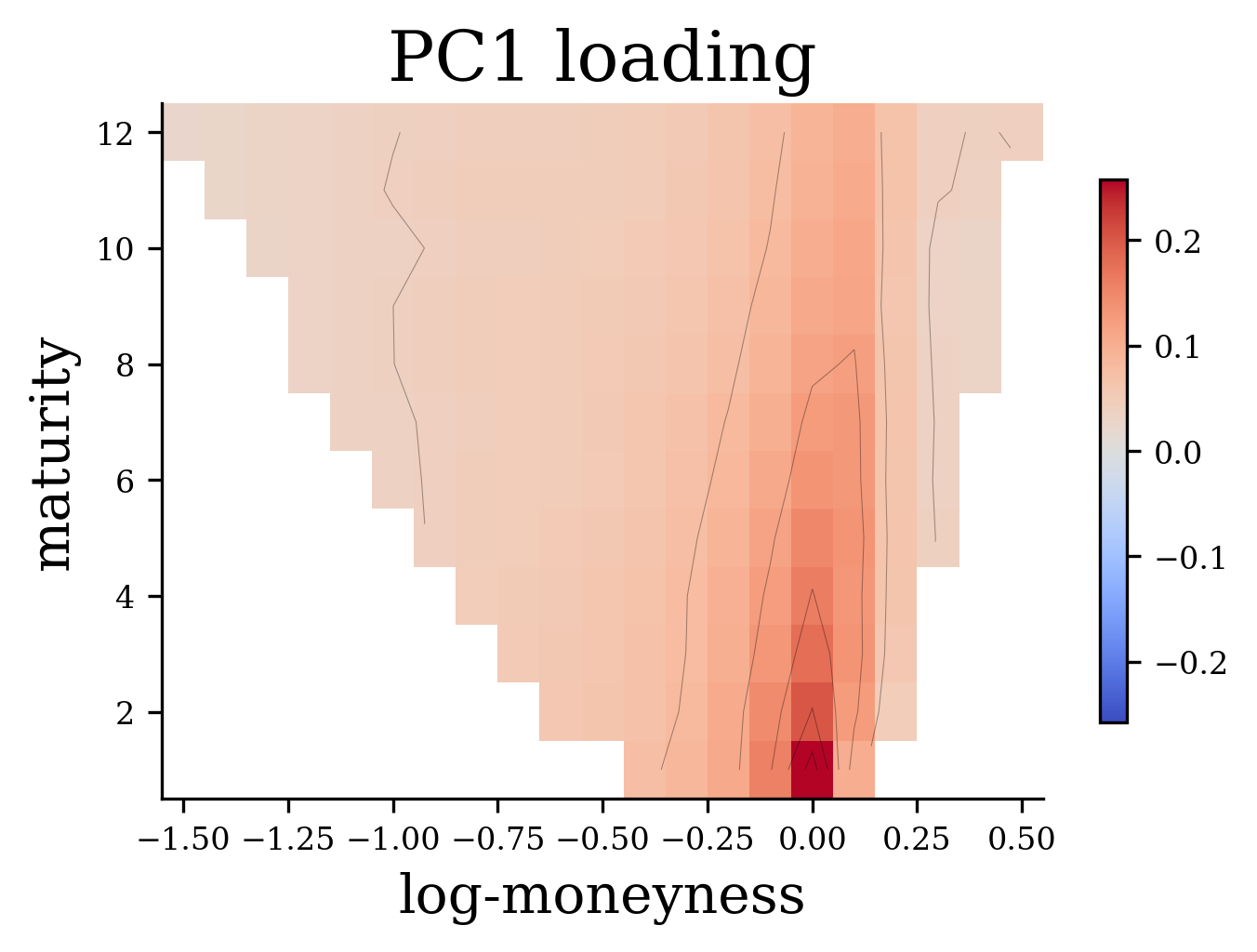}
    \includegraphics[width=.32\columnwidth]{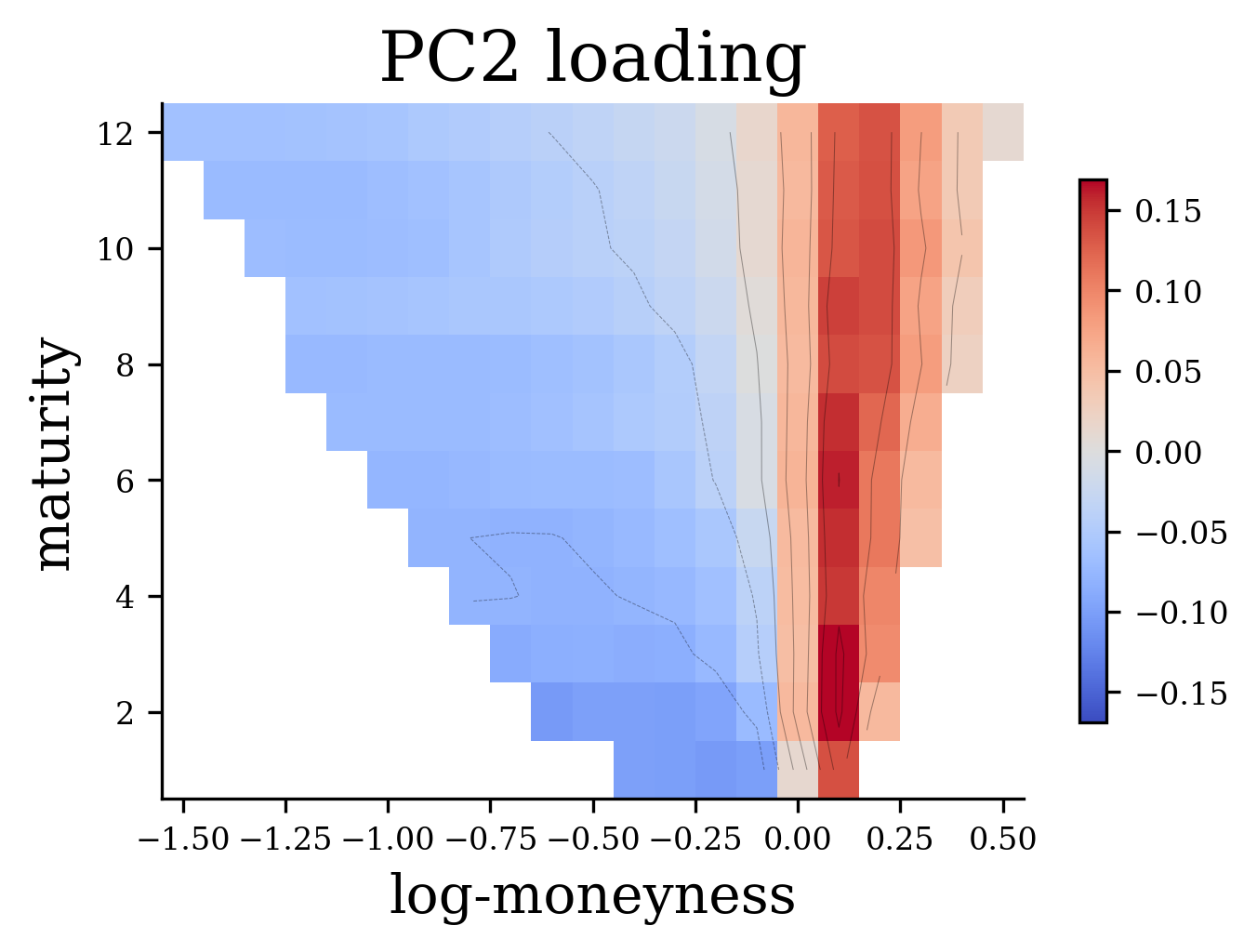}
    \includegraphics[width=.32\columnwidth]{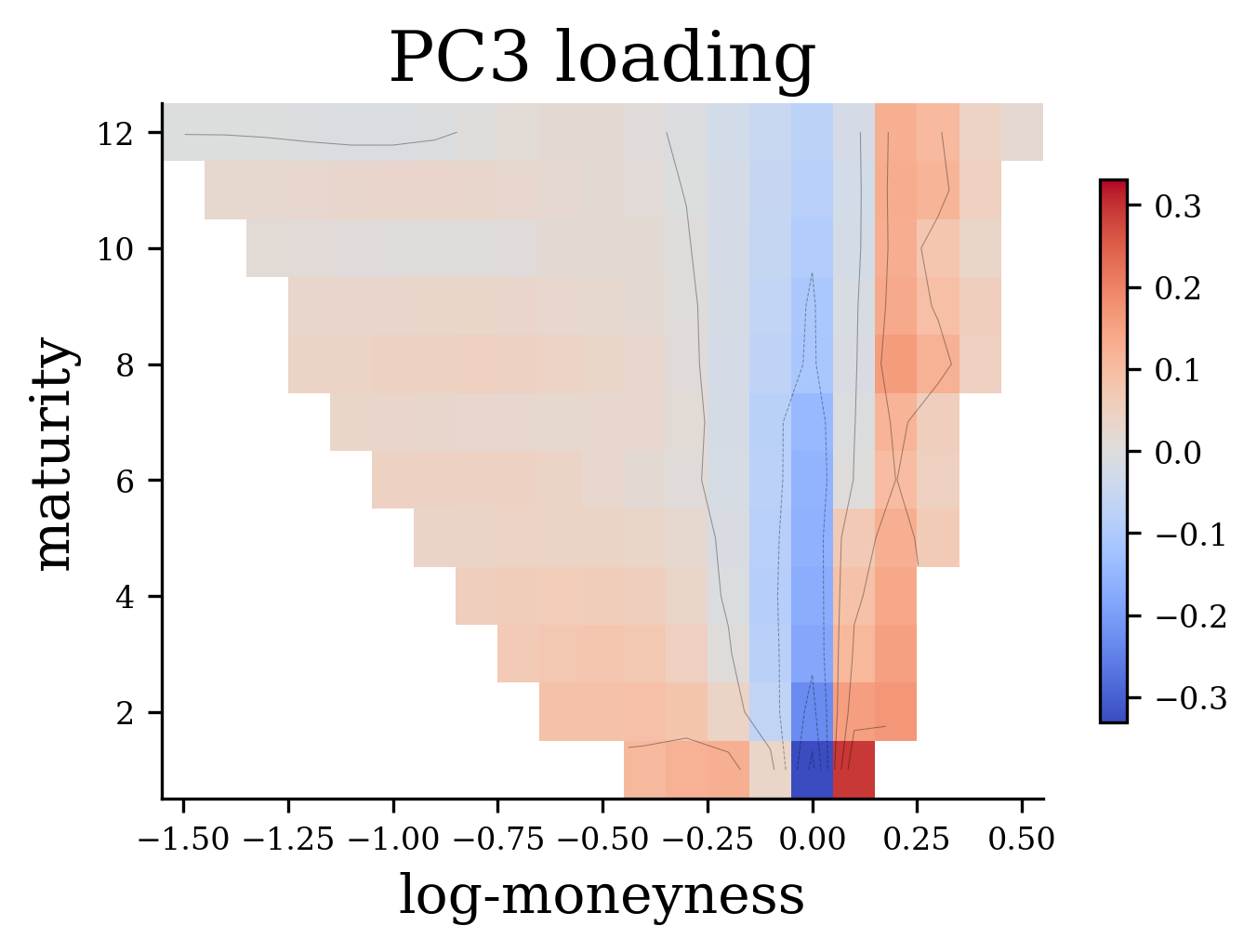}
    \Description{Three heat maps of principal-component loadings over log-moneyness and maturity. PC1 is predominantly positive across the surface, with its strongest loading near short maturities and approximately at-the-money coordinates, indicating a broad level movement. PC2 changes from negative loadings on the left wing to positive loadings near and above the centre, indicating a moneyness or skew tilt. PC3 alternates sign around the central moneyness region, with strong short-maturity variation, indicating a local curvature or smile-bending movement. Grey contour lines provide the average surface geometry for orientation.}
    \caption[Principal-component loadings of realised t--IVS increments]{Loadings of the first three principal components of realised log t--IVS increments.}
    \label{fig:sec:results:generative:pca-loadings}
\end{figure}

The loading surfaces in \autoref{fig:sec:results:generative:pca-loadings} provide clear economic interpretations. PC1 is broadly same-signed and captures level movements. PC2 changes sign across moneyness. PC3 bends around the central region. The model therefore reproduces both the concentration of variance in a few factors and the principal directions of surface deformation. Descriptor correlations at \(h=21\), reported in \autoref{tab:sec:results:generative:pca-interpretation}, strengthen these interpretations. PC1 is strongly associated with average-IV changes, PC2 with skew, and PC3 with curvature.

The main discrepancy concerns amplitude. Generated factor dispersion is close to realised dispersion at \(h=1\), but becomes increasingly understated at longer horizons. For PC1, generated and realised standard deviations are \(0.679\) and \(0.923\) at \(h=21\), and \(0.796\) and \(1.021\) at \(h=30\). Thus, generated surface movements align with the dominant empirical modes of variation, although their dispersion is increasingly understated at longer horizons.

\begin{figure}[!htpb]
    \centering
    \includegraphics[width=\columnwidth]{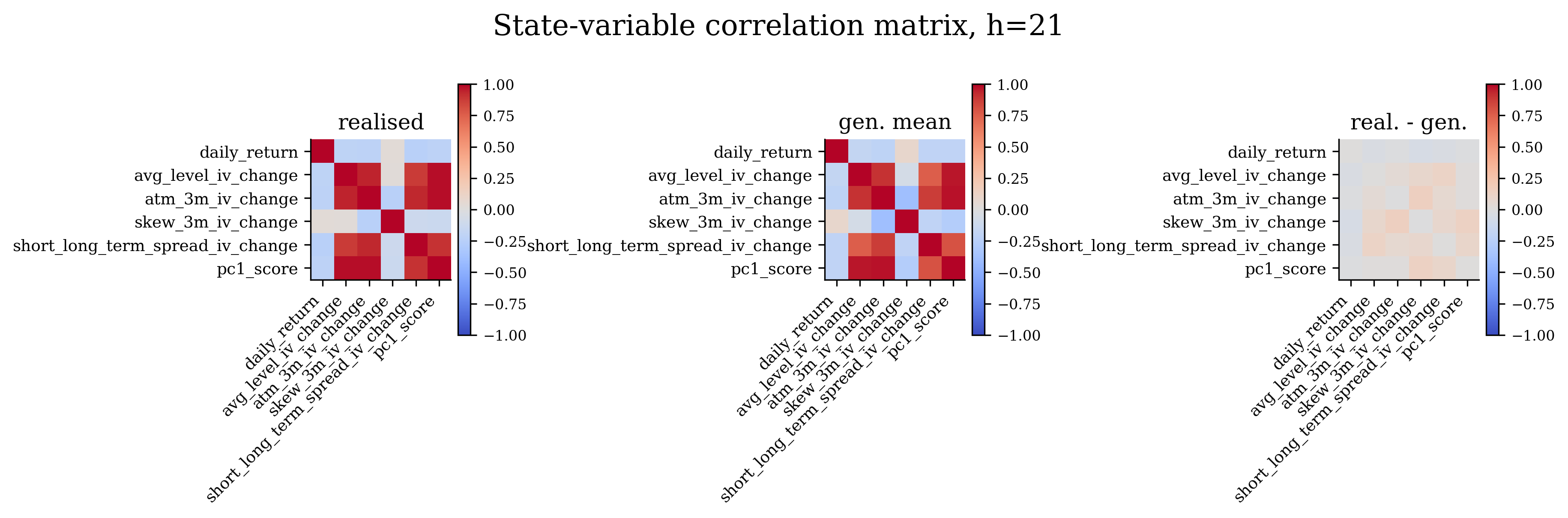}
    \Description{Three six-by-six heat maps at forecast horizon 21 comparing correlations among daily return, average implied volatility change, three-month at-the-money implied volatility change, three-month skew change, short-minus-long term-spread change, and the first principal-component score. The realised and generated matrices show similar blocks of strong positive dependence among the level-related surface variables and PC1, and negative dependence between returns and those variables. The realised-minus-generated matrix is mostly pale and close to zero, with only small residual differences. Red denotes positive correlation, blue negative correlation, and white values near zero.}
    \caption[State-variable dependence at (h=21)]{Realised correlations, mean generated correlations, and their difference at forecast horizon (h=21).}
    \label{fig:sec:results:generative:state-corr-h21}
\end{figure}

\paragraph{Joint dependence and return--surface dynamics}
\autoref{fig:sec:results:generative:state-corr-h21} compares realised and generated dependence among economically interpretable state variables at \(h=21\). The difference matrix is relatively muted from \(h=5\) onwards, indicating that the broader dependence structure is captured well.

The largest discrepancies concern the magnitude of return--surface dependence. \autoref{tab:sec:results:generative:return-surface-correlations} shows that the model reproduces its negative sign throughout, but understates the immediate leverage relation. At \(h=1\), the realised and generated correlations differ substantially, but agreement improves at multi-day horizons. The model therefore captures the main negative return--surface dependence at weekly to monthly horizons, but its first-day magnitude remains too weak. Correlations between returns and PC2/PC3 are small and are therefore not interpreted as robust leverage factors.

\begin{table}[!htpb]
    \caption[Return--surface correlation dynamics]{Realised/generated correlations between daily returns and selected surface changes.}
    \label{tab:sec:results:generative:return-surface-correlations}
    \centering
    \scriptsize
    \setlength{\tabcolsep}{5pt}
    \begin{tabular}{lrrrr}
    \toprule
    Horizon \(h\) & \(\Delta\) Average IV & \(\Delta\) ATM 3M IV & \(\Delta\) Term-spread & PC1 score \\
    \midrule
    (1)
    & \(-0.791 / -0.538\)
    & \(-0.860 / -0.616\)
    & \(-0.800 / -0.481\)
    & \(-0.854 / -0.607\) \\
    \addlinespace
    (5)
    & \(-0.326 / -0.304\)
    & \(-0.349 / -0.366\)
    & \(-0.324 / -0.325\)
    & \(-0.351 / -0.344\) \\
    \addlinespace
    (10)
    & \(-0.276 / -0.239\)
    & \(-0.297 / -0.284\)
    & \(-0.301 / -0.268\)
    & \(-0.296 / -0.269\) \\
    \addlinespace
    (21)
    & \(-0.208 / -0.175\)
    & \(-0.217 / -0.209\)
    & \(-0.235 / -0.196\)
    & \(-0.215 / -0.201\) \\
    \addlinespace
    (30)
    & \(-0.173 / -0.151\)
    & \(-0.179 / -0.169\)
    & \(-0.194 / -0.164\)
    & \(-0.183 / -0.167\) \\
    \bottomrule
    \end{tabular}
\end{table}

\paragraph{Economic admissibility}
We evaluate arbitrage restrictions using finite differences on the grid. Since a surface is classified as non-arbitrage-free if it contains any local violation, the free-surface share is interpreted jointly with violation frequency and magnitude.

\begin{table}[!htpb]
    \caption{No-arbitrage diagnostics. ``Cal.''/``Bfly.'' report violated checks; ``Repair'' reports the correction at each surface's worst grid point relative to the median half bid--ask spread.}
    \label{tab:sec:results:generative:no-arbitrage}
    \centering
    \footnotesize
    \setlength{\tabcolsep}{6pt}
    \begin{tabular}{lrrrrr}
    \toprule
    Surface & Free & Cal. & Bfly. & Mean repair & Max repair \\
    \midrule
    DDPM samples        & 88.5\% & 0.00\% & 0.13\% & 15\% & 240\% \\
    Ensemble mean       & 98.8\% & 0.00\% & 0.01\% & 7\%  & 24\%  \\
    Ensemble median     & 98.4\% & 0.00\% & 0.01\% & 7\%  & 22\%  \\
    \bottomrule
    \end{tabular}
\end{table}

\autoref{tab:sec:results:generative:no-arbitrage} shows virtually no calendar violations and only sparse butterfly violations. Detected violations are economically small: the required IV repair averages only \(15\%\) of the median half bid--ask spread for individual samples and \(7\%\) for the ensemble surfaces.\footnote{The repair is a local proxy: for each violated finite-difference constraint, we compute the minimum RMS adjustment to the involved ordinary-IV values required to reach the no-arbitrage region. Bid--ask half-spreads are obtained by applying operator deep smoothing separately to bid and ask quotes.} In absolute terms, the typical repair of \(1.4\) IV bps remains below local half-spreads across the grid, which range from about \(3.5\) bps at the money to \(45\) bps in the deep wings. Only rare tail violations in individual samples exceed the half-spread at their worst grid point. These results suggest that detected arbitrage largely reflects quote noise, non-synchronous observations, smoothing/interpolation artefacts, or other underlying causes, rather than economically material arbitrage.

Disabling the no-arbitrage mechanism (\(\lambda_{\mathrm{but}}=0, \lambda_{\mathrm{cal}}=0\)) cuts the arbitrage-free share to about \(25\%\) while corrections increase two- to fivefold (mostly still within the half-spread). This is most likely because the training surfaces themselves contain arbitrage only at small magnitudes, so the mechanism primarily ensures admissibility rather than preventing exploitable arbitrage.

\subsection{Distributional Forecast Quality}
\label{sec:results:distributional-quality}

\paragraph{Proper scoring rule}
We evaluate whether the conditional samples assign probability mass effectively to the realised future, using the continuous ranked probability score (CRPS) \cite{Gneiting07}. Deterministic persistence corresponds to a degenerate distribution. To obtain a non-degenerate comparison, we construct a historical-change bootstrap benchmark by sampling \(1000\) \(30\)-day surface-change trajectories from the training and validation period and anchoring them at each forecast origin. For returns, the benchmark is zero.

\autoref{tab:sec:results:distributional:crps} shows that the large gains relative to deterministic persistence partly reflect the benefit of accounting for predictive uncertainty. At \(h=1\), diffusion only modestly improves on bootstrap persistence for average and ATM levels and is worse for skew, curvature, and term spread. The comparison becomes favourable at longer horizons: by \(h=10\), diffusion outperforms bootstrap persistence in nine (of nine) surface descriptors and in seven (of nine) grid buckets; by \(h=30\), it outperforms bootstrap persistence in nine descriptors and eight grid buckets, with the only disadvantage in the central long-maturity region. Thus, deterministic persistence overstates the apparent short-horizon probabilistic gain, whereas diffusion retains broad distributional improvements at medium and longer horizons.

\begin{table}[!htpb]
    \caption[Marginal distributional forecast accuracy]{CRPS gains relative to deterministic / bootstrap persistence benchmarks. Positive gains indicate lower CRPS.}
    \label{tab:sec:results:distributional:crps}
    \centering
    \footnotesize
    \begin{tabular}{lrrr}
    \toprule
    Quantity & h=1 & h=10 & h=30 \\
    \midrule
    Average IV level
    & +24.9/ +3.6\%
    & +31.8/+10.8\%
    & +32.8/+10.5\% \\
    ATM IV level
    & +23.3/ +1.7\%
    & +29.2/ +6.4\%
    & +30.0/ +4.7\% \\
    3M skew
    & +10.2/-21.1\%
    & +34.2/ +7.3\%
    & +30.8/ +3.2\% \\
    3M curvature
    & +5.7/-26.9\%
    & +31.8/ +4.8\%
    & +29.2/ +1.9\% \\
    Short--long spread
    & -23.7/-57.4\%
    & +26.8/ +5.7\%
    & +35.5/+14.8\% \\
    Daily return
    & +27.4/\text{--}
    & +26.0/\text{--}
    & +25.8/\text{--} \\
    \bottomrule
    \end{tabular}
\end{table}

The multivariate energy score \cite{Gneiting07} exhibits a similar horizon-dependent pattern, with persistence performing better at one day and diffusion achieving lower scores at weekly-to-monthly horizons. The same results also hold when returns are included jointly.

\begin{figure}[!htpb]
    \centering
    \begin{minipage}[t]{0.48\columnwidth}
    \includegraphics[width=\columnwidth]{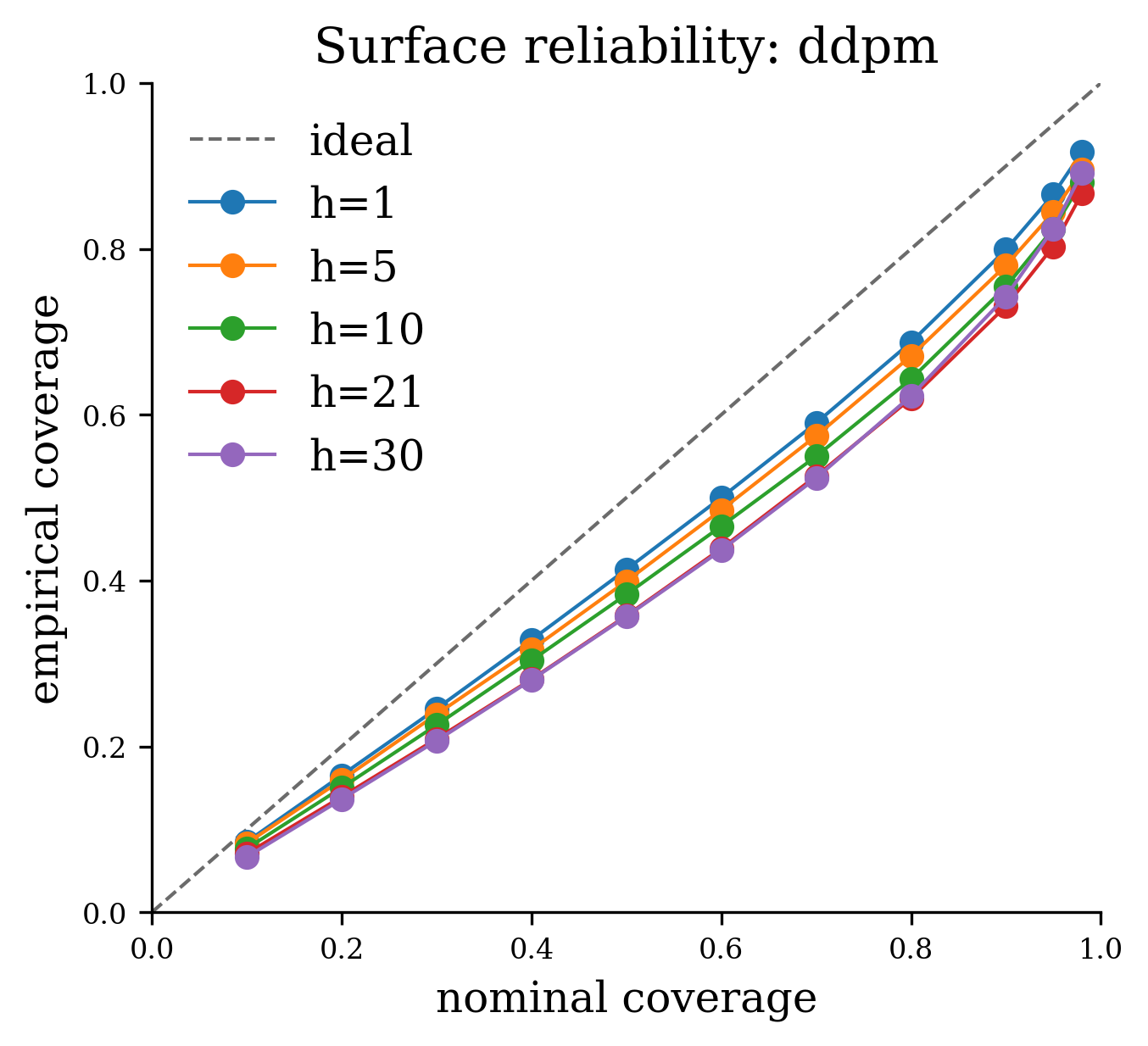}
    \end{minipage}
    \hfill
    \begin{minipage}[t]{0.48\columnwidth}
    \includegraphics[width=\columnwidth]{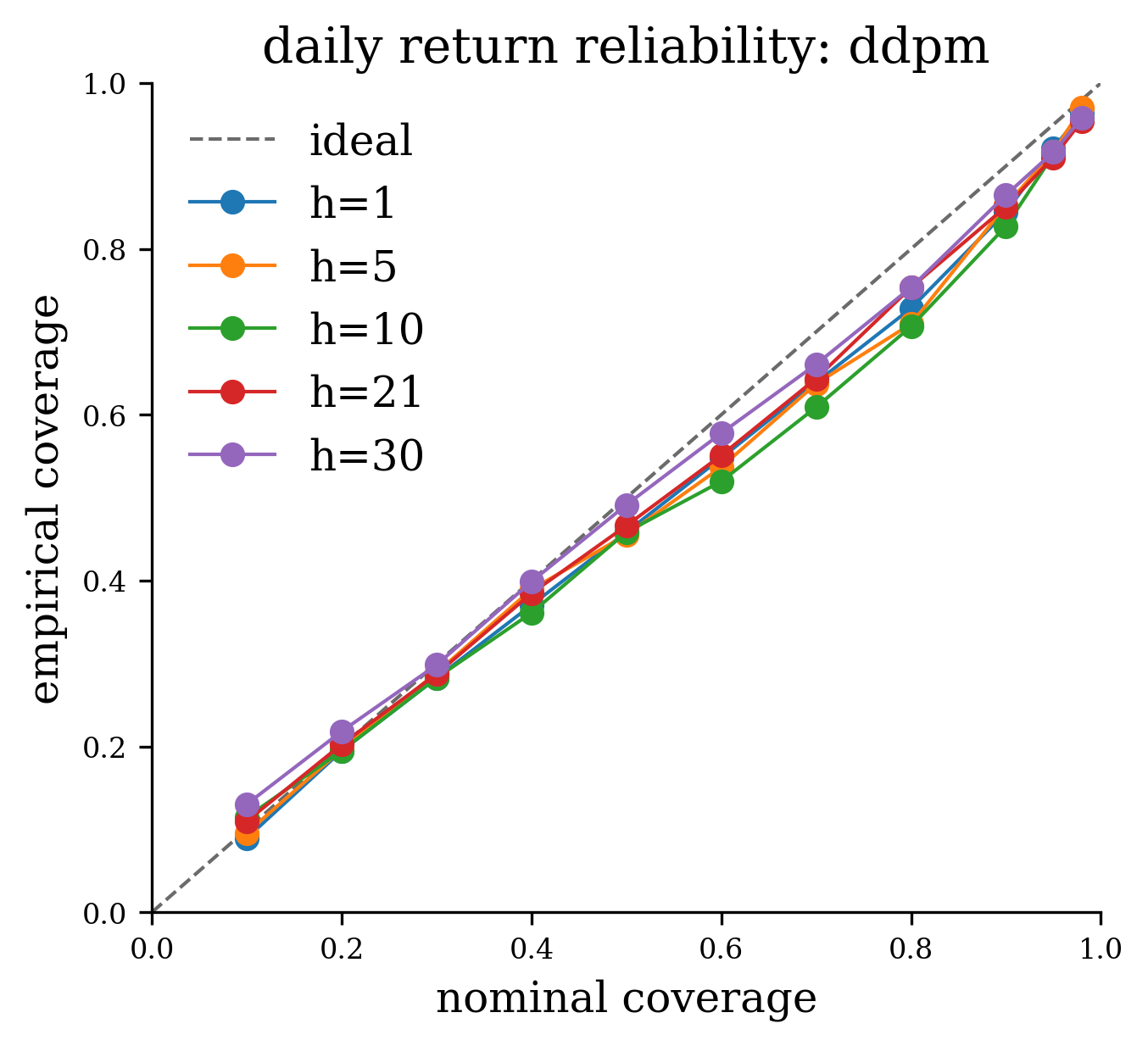}
    \end{minipage}
    \Description{Two reliability diagrams with nominal predictive-interval coverage on the horizontal axis and empirical coverage on the vertical axis. The left panel evaluates pointwise surface values and the right panel daily returns. A dashed diagonal represents perfect calibration, while separate coloured curves represent horizons 1, 5, 10, 21, and 30. All surface curves lie below the diagonal, with longer horizons generally farther below it, indicating undercoverage. Return curves are closer to the diagonal but also show modest undercoverage, especially at high nominal coverage.}
    \caption[Predictive-interval reliability]{Predictive-interval reliability for pointwise surface values and daily returns.}
    \label{fig:sec:results:distributional:reliability}
\end{figure}

\paragraph{Calibration}
We construct pointwise predictive intervals from the generated paths and average coverage over grid points and origins. \autoref{fig:sec:results:distributional:reliability} shows systematic surface undercoverage across horizons and interval levels. Coverage falls from \(0.413\) to \(0.357\) for nominal \(50\%\) intervals and from \(0.866\) to \(0.824\) for nominal \(95\%\) intervals between \(h=1\) and \(h=30\), despite monotonically increasing interval widths. The model therefore expands uncertainty over time, but not sufficiently. The gap narrows for the highest confidence levels. Moreover, calibration errors are heterogeneous. At \(h=1\), the largest undercoverage occurs in the short-maturity right wing, which remains the most affected at longer horizons. However, undercoverage becomes increasingly pronounced in the central region at medium-long maturities. These results are consistent with the factor diagnostics: the generated trajectories reproduce important empirical directions and dependence patterns, but understate the dispersion of future surface movements. Its samples are therefore informative but underdispersed, warranting calibration before applications such as risk management or stress testing.

Daily-return intervals lie closer to the diagonal, although their widest intervals also show slight undercoverage.

\subsection{Predictive Content Beyond Persistence}
\label{sec:results:predictive-content}

Finally, we ask whether the predictive mean also improves upon persistence as a point forecast. Because (SPX) options trade in a highly efficient market, persistence is a particularly demanding baseline: most available information is already reflected in the surface, so improvements of even a few percentage points can represent meaningful predictive gains. Conversely, not confronting persistence can flatter a model more than it informs the reader.

\paragraph{Surface improvement}
\autoref{tab:sec:results:predictive:surface-accuracy} shows a clear horizon transition. Aggregated over all horizons, the model reduces RMSE and MAE. Its aggregate signed bias remains small, suggesting that gains arise from horizon-dependent shape adjustments rather than systematic level shifts. At \(h=1\), persistence is substantially better: the model introduces too much movement. One possible explanation is that the model observes a \(21\)-day evolution and uses it to infer a continuation. Such recent dynamics may contain information about more persistent medium-horizon movements, while the next-day innovation itself remains noisy and difficult to predict. Part of the short-horizon difficulty also arises from the latent representation: AE-only reconstruction RMSE is \(0.0039\), which constitutes a comparatively larger share of total error at short horizons. Estimated gains turn positive from approximately \(h=4\) for RMSE and \(h=6\) for MAE.

These predictive improvements cannot be explained by autoencoder smoothing alone, which is shown in \autoref{tab:sec:results:predictive:ae-persistence}. The model mean frequently outperforms AE persistence, that is persistence after encoding and decoding, with only marginal gains at the shortest horizon and more pronounced improvements thereafter.

\begin{table}[!htpb]
    \caption[Surface point-forecast accuracy I]{Accuracy of the model mean relative to persistence.}
    \label{tab:sec:results:predictive:surface-accuracy}
    \centering
    \scriptsize
    \setlength{\tabcolsep}{4pt}
    \begin{tabular}{lrrrrr}
    \toprule
    Scope & RMSE & Persistence RMSE & RMSE gain & MAE gain & Bias \\
    \midrule
    All 30 & 0.01262 & 0.01343 & +6.09\% & +3.45\% & +0.00022 \\
    \midrule
    \(h=1\)  & 0.00619 & 0.00535 & -15.72\% & -34.49\% & +0.00002 \\
    \(h=5\)  & 0.01017 & 0.01043 & +2.52\%  & -0.75\%  & +0.00018 \\
    \(h=10\) & 0.01223 & 0.01305 & +6.28\%  & +4.46\%  & -0.00007 \\
    \(h=21\) & 0.01367 & 0.01459 & +6.29\%  & +4.30\%  & +0.00022 \\
    \(h=30\) & 0.01460 & 0.01567 & +6.83\%  & +4.95\%  & +0.00084 \\  
    \bottomrule
    \end{tabular}
\end{table}

\begin{table}[!htbp]
    \caption[Surface point-forecast accuracy II]{Accuracy of the model mean relative to AE persistence.}
    \label{tab:sec:results:predictive:ae-persistence}
    \centering
    \scriptsize
    \setlength{\tabcolsep}{4pt}
    \begin{tabular}{lrrrr}
    \toprule
    Scope & RMSE & AE-pers. RMSE & RMSE gain & MAE gain \\
    \midrule
    All 30 & 0.01262 & 0.01345 & +6.21\% & +5.09\% \\
    \midrule
    $h=1$  & 0.00619 & 0.00624 & +0.80\% & +0.07\% \\
    $h=5$  & 0.01017 & 0.01059 & +3.93\% & +3.10\% \\
    $h=10$ & 0.01223 & 0.01307 & +6.43\% & +5.57\% \\
    $h=21$ & 0.01367 & 0.01455 & +6.03\% & +5.10\% \\
    $h=30$ & 0.01460 & 0.01560 & +6.38\% & +4.77\% \\
    \bottomrule
    \end{tabular}
\end{table}

The predictive improvements also persist without the post-hoc mean-scaling gate: the ungated model\footnote{Both evaluations use the same trained autoencoder and diffusion models, but independently sampled trajectories (due to memory reasons). Performance differences may therefore also reflect Monte Carlo variability.} shown in \autoref{tab:sec:results:predictive:gate-ablation} yields positive RMSE gains at \(27\) of \(30\) horizons and improves aggregate RMSE and MAE relative to persistence by \(4.77\)\% and \(2.08\)\%, respectively.

\begin{table}[!htbp]
    \caption[Mean-scaling gate ablation]{Gated/ungated RMSE gains relative to persistence.}
    \label{tab:sec:results:predictive:gate-ablation}
    \centering
    \scriptsize
    \setlength{\tabcolsep}{4pt}
    \begin{tabular}{lrrrrr}
    \toprule
    Model & $h=1$ & $h=5$ & $h=10$ & $h=21$ & $h=30$ \\
    \midrule
    Gated   & -15.72\% & +2.52\% & +6.28\% & +6.29\% & +6.83\% \\
    Ungated & -19.06\% & +1.79\% & +6.61\% & +3.44\% & +4.83\% \\
    \bottomrule
    \end{tabular}
\end{table}

\begin{figure}[!htpb]
    \centering
    \includegraphics[width=\columnwidth]{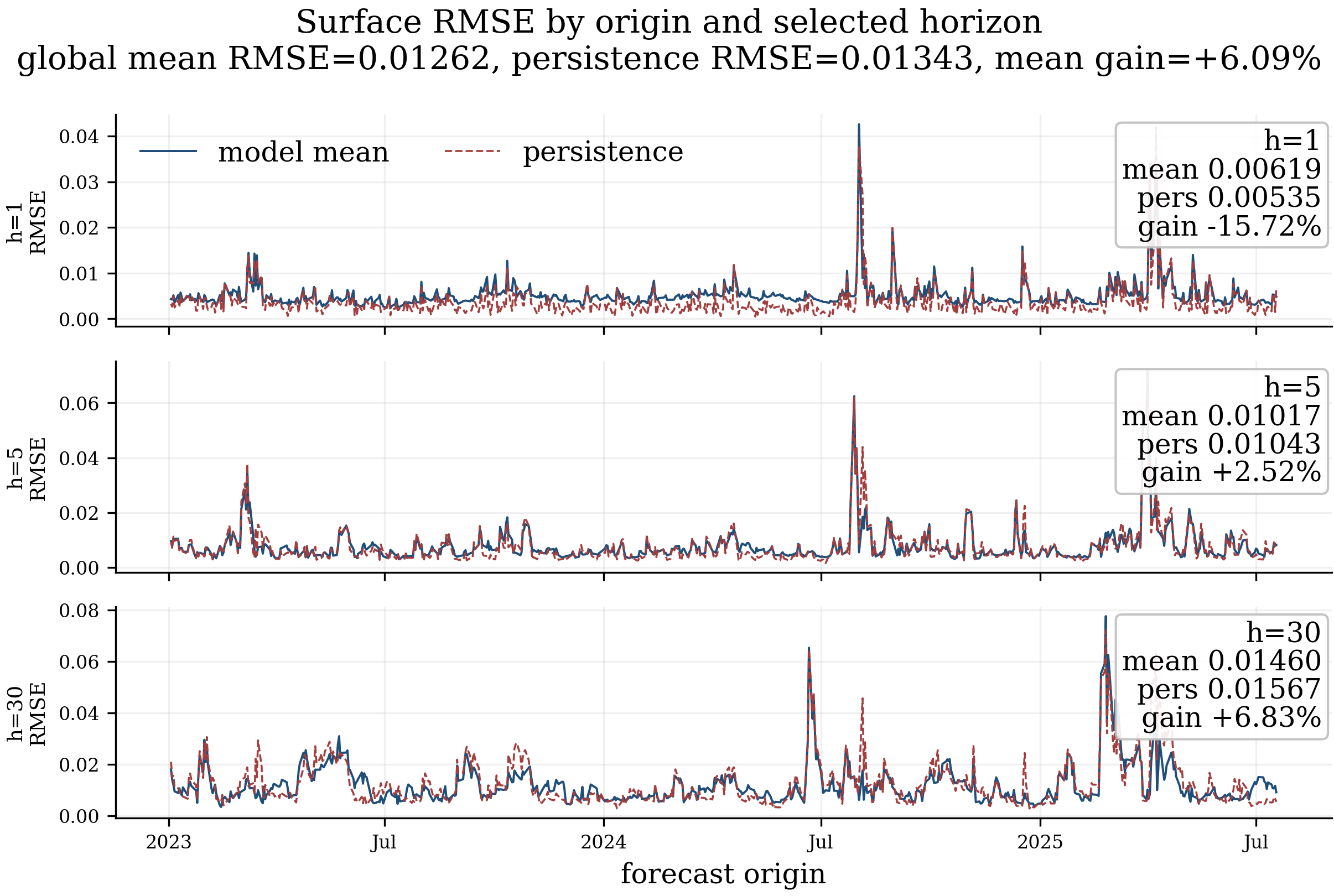}
    \Description{Three vertically stacked time-series panels comparing surface root-mean-square error for the model mean and persistence across forecast origins from 2023 to mid-2025. Panels correspond to horizons 1, 5, and 30 days. The model is shown by a solid blue line and persistence by a dashed red line. Both methods experience error spikes during the same difficult periods. At one day the model is generally above persistence, while at five and especially thirty days it is frequently below persistence. Text boxes report average error and percentage gain for each horizon.}
    \caption[Surface RMSE over time by forecast horizon]{Surface RMSE of the model mean and persistence across test origins at selected horizons.}
    \label{fig:sec:results:predictive:surface-rmse-time}
\end{figure}

\autoref{fig:sec:results:predictive:surface-rmse-time} shows that the gains are not driven by a few origins. Model and persistence errors co-move strongly and spike during the same difficult market episodes. The model therefore does not eliminate inherently difficult episodes but produces recurrent incremental gains once persistence becomes less dominant.

Moreover, performance is heterogeneous across the surface. Aggregated over all horizons, RMSE improves by \(11.69\%\) at short maturities, \(6.46\%\) at medium maturities, and \(4.99\%\) at long maturities. Both wings improve by approximately \(6\%\)--\(7\%\), whereas the central region improves by only \(2.45\%\) in RMSE. From \(h=10\) onwards, all maturity--moneyness buckets except the long-maturity central region improve upon persistence. Gains are therefore strongest where the surface moves most and weakest in the stable central region with little movement for the model to exploit.

\begin{figure}[!htpb]
    \centering
    \includegraphics[width=\columnwidth]{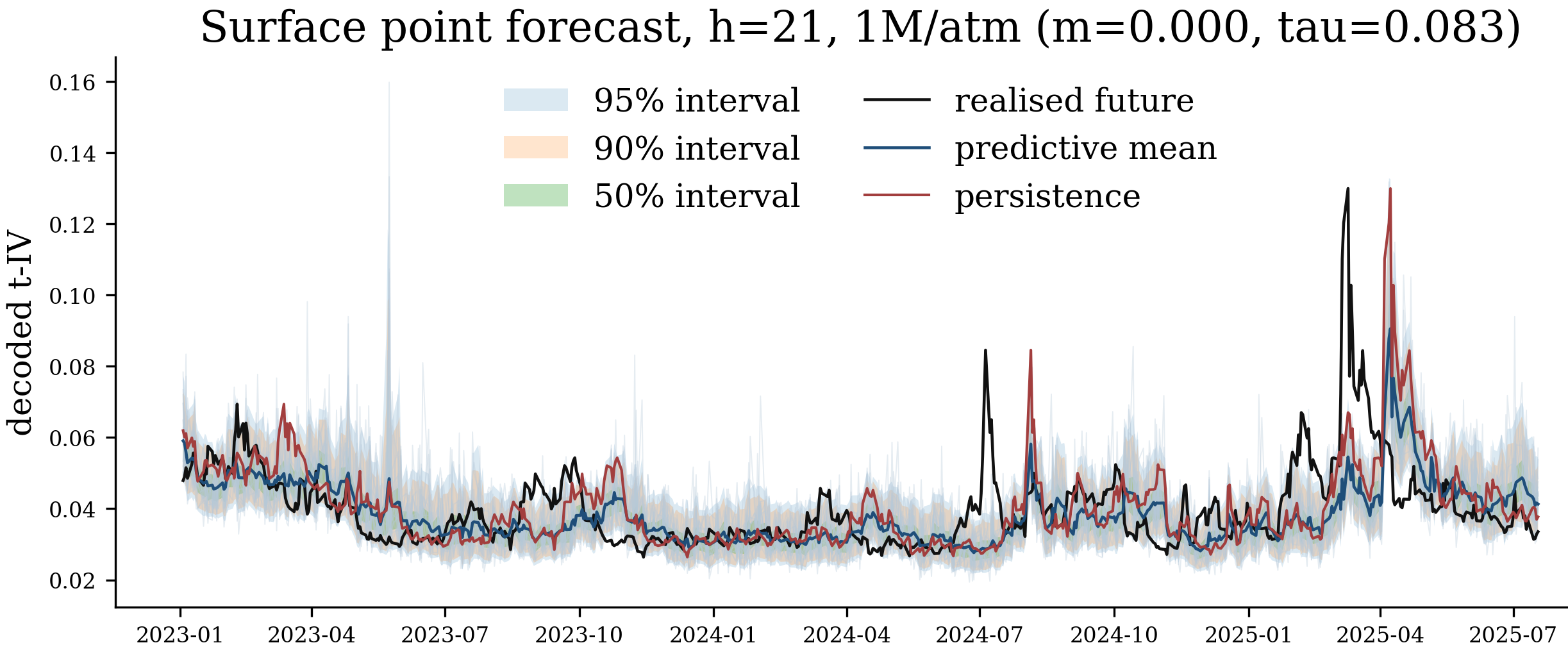}
    \Description{A time-series plot of decoded time-scaled implied volatility for the one-month at-the-money coordinate at horizon 21, shown across forecast origins from 2023 to mid-2025. The black realised series, blue predictive mean, and red persistence forecast generally move together, with pronounced volatility spikes in mid-2024 and early 2025. Nested green, orange, and light-blue areas show 50, 90, and 95 percent predictive intervals, and many faint sample paths show the forecast distribution. The predictive mean is smoother than the realised series and persistence, while interval widths increase around volatile periods.}
    \caption[Illustrative surface point forecast]{Realised future t--IV, model predictive mean, persistence, predictive intervals, and sampled forecast paths.}
    \label{fig:sec:results:predictive:representative-surface-points}
\end{figure}

The example point forecast in \autoref{fig:sec:results:predictive:representative-surface-points} illustrates this trade-off. Small departures from persistence are costly when the future surface barely moves, particularly at short horizons. At longer horizons, the predictive mean makes smooth, locally plausible adjustments while the generated paths remain coherent. The model's gains therefore arise from modest departures from persistence rather than unstable point predictions.

\paragraph{Return channel}
The return channel shows little point-forecast advantage over the zero-return baseline and is best interpreted as part of the joint trajectory generation: it produces informative distributions and realistic return--surface dependence without implying directional return predictability.

\section{Conclusion}
\label{cp:conclusion}

We introduced a conditional latent diffusion framework for generating \(30\)-day joint trajectories of implied volatility surfaces and returns. The model produces realistic, coherent, low-dimensional, and essentially arbitrage-free scenarios, with point forecast improvements over persistence emerging primarily at multi-day horizons. Its main limitations are underdispersion and weak one-day-ahead point accuracy. The framework is therefore best viewed as a probabilistic multi-step forecaster and conditional scenario generator, rather than a one-day point forecaster.

Future research could explore alternative arbitrage-free representations, such as DYSANOS \cite{BuehlerEtAl26DYSANOS} and SANOS-based \cite{BuehlerEtAl26SANOS} diffusion, alongside other generative approaches, including latent flow matching \cite{BrooksBajalica26} and random convolutional feature matching \cite{MuellerZozoulenko26}. Evaluating these approaches across stress regimes and less liquid underlyings would further establish their generalisability. More fundamentally, future work should identify which metrics are appropriate for evaluating functional trajectories.

The practical value of probabilistic IVS forecasts should ultimately be assessed in downstream applications such as hedging, risk measurement, and stress testing. In particular, the joint generation of IVS and return trajectories provides a natural basis for scenario-driven hedging approaches, including deep hedging under market frictions \cite{BuehlerGonon19, BuehlerEtAl21, MuellerAkkari24} and recent methods that use generative IVS models to construct and evaluate hedging strategies \cite{ContVuletic25, FrancoisEtAl25}. Such applications would provide an economically meaningful complement to statistical forecast evaluation, testing whether improvements in probabilistic trajectory modelling translate into better financial decisions.

\bibliographystyle{ACM-Reference-Format}
\bibliography{bibfile}

\end{document}